\documentclass[st,letterpaper,twocolumn]{jpsj3}
\usepackage{txfonts}
\usepackage[utf8]{inputenc}
\usepackage{amssymb}
\usepackage{amsmath}
\usepackage{graphics, graphicx}
\usepackage{mathcomp}
\usepackage{float}
\usepackage{sansmath}
\usepackage{color}

\title{Novel Approaches to Spectral Properties of Correlated
Electron Materials:\\
From Generalized Kohn-Sham Theory to Screened Exchange
Dynamical Mean Field Theory}

\author{Pascal Delange$^1$\thanks{pascal.delange@polytechnique.edu}, 
Steffen Backes$^1$\thanks{steffen.backes@polytechnique.edu}, 
Ambroise van Roekeghem$^2$\thanks{ambroise.vanroekeghem@gmail.com}, 
Leonid Pourovskii$^{1,3}$\thanks{leonid@cpht.polytechnique.fr},
Hong Jiang$^4$\thanks{jianghchem@pku.edu.cn}, 
and 
Silke Biermann$^{1,3,5}$\thanks{biermann@cpht.polytechnique.fr}}

\inst{$^1$Centre de Physique Th\'eorique, \'Ecole Polytechnique,
91128 Palaiseau, France
\\
$^2$CEA, LITEN, 17 Rue des Martyrs, 38054 Grenoble, France
\\
$^3$Coll\`ege de France, 11 place Marcelin Berthelot, 75005 Paris, France
\\
$^4$College of Chemistry and Molecular Engineering, Peking University, 100871 Beijing, China
\\
$^5$European Theoretical Spectroscopy Facility (ETSF), Europe
} 

\abst{
The most intriguing properties of emergent materials are
typically consequences of highly correlated quantum states of their
electronic degrees of freedom. Describing those materials from first principles
remains a challenge for modern condensed matter theory.
Here, we review, apply and discuss novel approaches to spectral
properties of correlated electron materials, assessing current day
predictive capabilities of electronic structure calculations.
In particular, we focus on the recent Screened Exchange Dynamical
Mean-Field Theory scheme and its relation to generalized 
Kohn-Sham Theory. These concepts are illustrated on the 
transition metal pnictide BaCo$_2$As$_2$ and
elemental zinc and cadmium.
}

\def\bra{\langle}
\def\ket{\rangle}

\def\t{\mbox{tr}\,}

\begin{document}
\maketitle


\section{Introduction}

Technological progress has been intimately related with progress
in materials science since its very early days. 
The last century has seen the development of refined capabilities
for materials elaboration, characterisation and control of properties, 
culminating among others in the unprecedented possibilities of the
digital age: materials properties have become an object of 
theoretical {\it simulations}, stimulating systematic searches
for systems with desired characteristics.
This branch of condensed matter physics is continuously evolving 
into a true new pillar of materials science,
where the theoretical assessment of solid state systems becomes
a fundamental tool for materials screening.

Obviously, the success of this program hinges on the degree
of {\it predictive} power of modern simulation techniques,
which have to allow for performing calculations without introducing
adjustable parameters. This requirement is even more severe since
the most popular branch of {\it ab initio} 
calculations, the Density Functional Theory (DFT) approach, is restricted to ground
state properties of the solid, while most properties of potential 
technological interest stem from excited states: the calculation
of any type of transport phenomenon, for example electric, thermal, 
thermoelectric or magnetoelectric transport, of optical or spectroscopic 
properties, or of magnetic, charge- or orbital susceptibilites
requires the accurate assessment of highly non-trivial response 
functions within a finite-temperature description.

It is not a coincidence that the materials with the most exotic
features are also most challenging to get to grips with from
a theoretical point of view: both the complexity of their
properties and the difficulty of their description are in fact
a consequence of the very nature of the underlying electronic
structure: typically, one is dealing with transition metal,
lanthanide or actinide compounds, where the electrons in partially
filled $d$- or $f$-shells strongly interact with each other through
electron-electron Coulomb interactions, leading to highly 
entangled many-body quantum states.

Even when using a simplified description of the
solid in terms of an effective lattice model, assessing these quantum
correlated states is a tremendous challenge. If {\it local}
quantum fluctuations on a given atomic site
are dominant, dynamical mean-field theory (DMFT) 
\cite{georges1996,kotliar_dmft_physicstoday}
yields an accurate description of the system.
In particular, DMFT captures both the strong coupling 
Mott-insulating limit and the weakly interacting band limit
of the Hubbard model, and is able to describe the salient
spectral features of a correlated metal with coexisting quasi-particle and Hubbard peaks
even in the intermediate correlation regime.
In the multi-orbital case, additional degrees of freedom
can lead to even richer physics with unconventional 
(e.g. orbital-selective) behavior \cite{PhysRevLett.95.206401}, 
or complex ordering phenomena \cite{PavariniPRL92-04,
MartinsPRL107-11,AritaPRL108-12,AhnKunesJPCM27-15,
PhysRevLett.106.256401,PhysRevB.90.235140}.

In order to recover the material-specific character of the
calculations, DMFT has been combined with DFT 
\cite{anisimov1997,lichtenstein1998} into the so-called
``DFT+DMFT'' scheme, which is nowadays one of the most
popular workhorses of electronic structure theory for correlated
electron materials.
Successful applications include transition metals
\cite{Fe, PhysRevB.94.100102,PhysRevB.82.104414}, transition metal oxides
\cite{PavariniPRL92-04,biermann:026404,PoteryaevPRB76-07,
tomczak_vo2_proc, optic_epl, tomczak_v2o3_proc,arpes_review,
Nekrasov2013, Thunstrom, PhysRevB.94.155135,AnisimovPRB71-05,PhysRevLett.93.156402,
mo_V2O3_prominent_peak}, 
lanthanide \cite{PhysRevB.76.235101,PhysRevB.94.085137} or 
actinide \cite{deltaPu,PhysRevB.72.115106,PhysRevB.92.085125,
PhysRevB.75.235107} 
systems.
Comparisons between calculated and measured spectral functions
have sometimes led to impressive agreement\cite{PhysRevLett.113.266407,PhysRevB.92.195128}.
However, despite these successes the construction of the 
Hamiltonian used in these calculations is rather {\it ad hoc}
and remains a limitation to the predictive power of the approach.
Therefore, the elaboration of
more systematic interfaces between electronic structure and
many-body theory has become an active area of research
\cite{hirayama2015}.

In this work, we review a recent scheme combining screened
exchange and DMFT \cite{PhysRevLett.113.266403,
vanRoekeghem-CaFe2As2-PRB2016,0295-5075-108-5-57003}.
We discuss its relation to many-body
perturbation theory and generalized Kohn-Sham (KS) Theory
and analyse the effects included at
the different levels of the theory. As an illustration we describe
its application to the low-energy spectral properties of the 
cobalt pnictide BaCo$_2$As$_2$.

How does a theory thas has been designed 
for strongly correlated materials 
reduce to an {\it a priori} simpler version in the case of
a weakly correlated system ?
For methods based on DFT that include only the static on-site Coulomb interaction
between localized states, like DFT+U\cite{anisimov_lda+u_review_1997_jpcm} or
DFT+DMFT, the spectrum in the limit of vanishing Hubbard interactions reduces
trivially to the Kohn-Sham spectrum of DFT.
For more complex interfaces of electronic structure and many-body
theory which we will discuss here, however,
it is a non-trivial question and allows for
interesting possibilities to check their consistency.
Here, we will discuss this issue on the example of the transition
metals zinc and cadmium. This also allows us to comment on the 
challenge of including states in a wider energy range, identifying a 
challenging problem on these seemingly ``simple'' systems. 

The paper is organised as follows: in section \ref{sec:gwdmft_to_sexdmft} we 
review how Screened Exchange Dynamical Mean Field Theory derives
from the combined many-body perturbation theory
+ dynamical mean field scheme ``GW+DMFT''.
In section \ref{sec:generalized_dft} we analyse the relation of screened
exchange schemes to generalized KS theory. 
Section \ref{sec:baco2as2} provides an example of the application of
such methods to BaCo$_2$As$_2$,
while section \ref{sec:Zn_Cd} discusses the electronic structure
of elemental zinc and cadmium within a simplified approach.
Finally, we conclude in section \ref{sec:conclusions}.


\section{From GW+DMFT to Screened Exchange Dynamical Mean Field
Theory}
\label{sec:gwdmft_to_sexdmft}
Improving the predictive power of methods that combine electronic structure
and many-body theory poses the challenge of properly connecting
the two worlds, without double counting of interactions
or screening. 
At the heart of this challenge lies the mismatch between
the density-based description of DFT and the Green's function
formalism used at the many-body level, as well as the difficulty 
of incorporating the feedback of high-energy screening processes
governed by the unscreened Coulomb interaction onto the low-energy
electronic structure.
Conceptually speaking, these difficulties can be avoided by
working on a large energy scale in the continuum with the full 
long-range Coulomb interactions, and a Green's function-based 
formalism even at the level of the weakly correlated states.
These features are realised within the combined many-body perturbation
theory and dynamical mean field theory scheme ``GW+DMFT'' 
\cite{PhysRevLett.90.086402,Sun2004,TomczakEPL100-12,
GWDMFT_review,Ayral2012,Ayral2013,Hansmann2013, Huang2014,
Boehnke2016, Ayral2017}: 
screening is assessed by the random phase 
approximation in the continuum, augmented by a local vertex correction,
while the starting electronic structure for the DMFT calculation
can be roughly interpreted as a ``non-local GW'' calculation 
\cite{PhysRevB.90.165138}.

Screened Exchange Dynamical Mean Field Theory \cite{PhysRevLett.113.266403,
vanRoekeghem-CaFe2As2-PRB2016,0295-5075-108-5-57003} can be
understood as an approximation to this full GW+DMFT scheme.
It is based on the recent observation 
\cite{PhysRevLett.113.266403,PhysRevB.90.165138} 
that within the \textit{GW} approximation the
correction to LDA can be split into two contributions:
a local dynamical self-energy $\Sigma_{loc}(\omega)$ and a k-dependent
but static self-energy $\Sigma_{nloc}(k)$, which does not contain any
local component. If such a separation was strictly valid in the
full energy range, the non-local static part of the full self-energy would
be given by the non-local Hartree-Fock contribution, since the dynamical part
vanishes at high frequency. 
In many realistic systems 
such a decomposition holds to a good approximation in the low-energy
regime that we are interested in, where the static part $\Sigma_{nloc}(k)$
is quite different from the Fock exchange term. It is 
approximately given by a screened exchange self-energy, leading to 
a decomposition of the \textit{GW} self-energy into 
$\Sigma_{GW}=[GW(\nu=0)]_{nonloc}+[GW]_{loc}$.
Here, the first term is a screened exchange contribution arising from the 
screened interaction $W(\nu)$ evaluated at zero frequency $W(\nu=0)$. 
The second term is the local projection of the $GW$ self-energy. 
It is simply given by the \textit{GW} self-energy evaluated 
using a local propagator $G_{loc}$ and the local screened Coulomb interaction 
$W_{loc}$. Exactly as in GW+DMFT, 
Screened Exchange Dynamical Mean Field Theory replaces this term by
a non-perturbative one: it is calculated from an effective local impurity 
problem with dynamical interactions. 
In current practical 
applications of Screened Exchange + DMFT the RPA-screened Coulomb potential 
$W$ has been replaced by its long-wavelength limit, which reduces to a simple 
Yukawa-form~\cite{PhysRevLett.113.266403}.

Quite generally the dynamical character of the interactions results in an 
additional renormalization $Z_B$ of the hopping amplitudes, which can be 
understood as an \textit{electronic polaron} effect: 
the coupling of the electrons to plasmonic screening degrees of
freedom leads to an effective mass enhancement corresponding
to the hopping reduction, manifesting itself as a narrowing
of the band. This effect can be estimated 
from the plasmon density of modes as given by the imaginary part
of the frequency dependent interaction $W$.  An explicit expression
for $Z_B$ has been derived in Ref.~\citen{PhysRevLett.109.126408}.


\section{Relation to Generalized Kohn-Sham Theory }
\label{sec:generalized_dft}

Screened Exchange Dynamical Mean Field Theory can also be
viewed as a specific approximation to a spectral density functional
theory based on the Generalized Kohn-Sham (GKS)
scheme of Seidl \textit{et al.}\cite{seidl1996}.
In GKS theory, alternative choices for the reference system that are different
than the familiar Kohn-Sham system of DFT are explored.
In particular, a generalized Kohn-Sham scheme where the
reference system is a screened exchange Hamiltonian can be
constructed. The main motivation for the inclusion of 
screened exchange in the literature has been to improve
upon the band gap problem in semiconductors. Indeed, it can
be shown that the screened exchange contribution, which 
corresponds to a non-local potential, effectively reintroduces
to some degree the derivative discontinuity that is missing in the pure
DFT description based on local exchange-correlation 
potentials\cite{PhysRevLett.49.1691}. Since the derivative discontinuity corresponds
to the discrepancy between the true gap and the Kohn-Sham
gap in exact DFT, a substantial improvement of the theoretical estimate
for band gaps can be expected on physical grounds and has
indeed been found.
Here, our goal is somewhat different: motivated by the
analysis of the role of screened exchange in GW+DMFT described
above, we would like to connect the Screened Exchange DMFT
scheme introduced above to generalized KS schemes making direct use
of the non-local screened exchange potential.

With this aim in mind, we briefly review the generalized Kohn-Sham 
construction in the case an effective Kohn-Sham system
including screened exchange.
Hereby, we follow closely Seidl \textit{et al.}\cite{seidl1996},
both in  notation and presentation.
First, one defines a functional
\begin{align}
S\left[ \Phi \right] = \left\langle \Phi \left| \hat{T} \right| \Phi \right\rangle  + U_H\left[\{ \phi_i \}\right] + E^{sx}_x\left[\{ \phi_i \}\right]
\end{align}
that includes, in addition to the familiar kinetic energy
term $\langle \Phi | \hat{T} | \Phi \rangle$ and the Hartree energy
$U_H\left[\{ \phi_i \}\right]$ also the screened
Fock term 
\begin{align}
E^{sx}_x\left[\{ \phi_i \}\right] =& - \sum\limits_{i<j}^{N} \int d\boldsymbol{r}d\boldsymbol{r'}  \nonumber \\ 
		&\times \cfrac{\phi_i^*(\boldsymbol{r})\phi_j^*(\boldsymbol{r'})  e^{-k_{\mathrm{TF}}  \left| \boldsymbol{r} - \boldsymbol{r'} \right|}  \phi_j(\boldsymbol{r})\phi_i(\boldsymbol{r'})  }{ \left| \boldsymbol{r} - \boldsymbol{r'} \right|}
\end{align}
Here, $\Phi$ are Slater determinants of single-particle
states $ \phi_i $. $k_{\mathrm{TF}}$ is the Thomas Fermi wave vector.
In order to derive a functional of the density Seidl
\textit{et al.} define a functional $F^s$ via the minimisation
\begin{eqnarray}
F^S\left[ \rho \right] = \min_{\Phi \rightarrow \rho(\boldsymbol{r})} S\left[ \Phi \right] = \min_{\left\{ \phi_i\right\} \rightarrow \rho(\boldsymbol{r})} S\left[ \left\{ \phi_i\right\} \right]
\end{eqnarray}
Next we define the energy functional 
\begin{eqnarray}
E^S\left[ \{ \phi_i \}; v_{eff} \right] = S\left[ \{ \phi_i \} \right] + \int d\boldsymbol{r} v_{eff}(\boldsymbol{r}) \rho(\boldsymbol{r})
\end{eqnarray}
where now the potential $v_{eff}$ does not only include the
external potential $v$ as in usual DFT, but also a contribution
by the exchange-correlation part
\begin{eqnarray}
v_{eff} = v + v_{xc}^{sx}\left[ \rho \right].
\end{eqnarray}
The additional contribution, the generalized (local) exchange-correlation
potential 
\begin{eqnarray}
v_{xc}^{sx} = \frac{\partial E_{xc}^{sx}\left[\rho\right]}{\partial \rho}.
\end{eqnarray}
is the functional derivative of the generalized
(local) exchange-correlation functional
\begin{eqnarray}
 E_{xc}^{sx}\left[\rho\right]=
 E_{xc}\left[\rho\right]- E_{x}^{sx}\left[\rho\right]+
 T\left[\rho\right]- T^{sx}\left[\rho\right]
\end{eqnarray}
which comprises the difference between the exchange-correlation potential
of standard Kohn-Sham DFT and the non-local exchange energy defined
above, as well as the difference between the kinetic energies of the
standard and generalized Kohn-Sham systems.
The functional derivative will eventually have to be 
evaluated self-consistently at the converged density.

This construction leads to the generalized Kohn-Sham
equations
\begin{align}
- \nabla^2 &\phi_i(\boldsymbol{r}) + v(\boldsymbol{r}) \phi_i(\boldsymbol{r}) + u\left([\rho];\boldsymbol{r}\right) \phi_i(\boldsymbol{r})  \nonumber \\
 &- \int d\boldsymbol{r'} v_x^{sx,\mathrm{NL}}(\boldsymbol{r},\boldsymbol{r'})\phi_i(\boldsymbol{r'}) + v_{xc}^{sx}\left( [\rho];\boldsymbol{r} \right) \phi_i(\boldsymbol{r})  = \varepsilon_i \phi_i  
\end{align}
with the Hartree potential $u$ and the non-local screened Fock potential
\begin{align}
 v_{x}^{sx,\mathrm{NL}}(\boldsymbol{r},\boldsymbol{r'}) &= - \sum\limits_{j=1}^{N} \cfrac{\phi_i(\boldsymbol{r}) e^{-k_{\mathrm{TF}}  \left| \boldsymbol{r} - \boldsymbol{r'} \right|} \phi_j^*(\boldsymbol{r'}) }{\left| \boldsymbol{r} - \boldsymbol{r'} \right|}.
\end{align}
and the effective (local) generalized Kohn Sham potential
$v_{xc}^{sx} $ defined above.
The generalized Kohn-Sham equations have the form
\begin{eqnarray}
\hat{\mathcal{O}}
\left[ \{ \phi_i \} \right]
\phi_j + \hat{v}_{eff} \phi_j = \epsilon_j \phi_j
\end{eqnarray}
where 
$\hat{\mathcal{O}}$ is a non-local operator, generalizing
the standard Kohn-Sham operator consisting solely of kinetic
energy and Hartree potential. 

The ground state energy for a system in the external potential
$v$ is then given by the expression
\begin{eqnarray}
E_{{\rm SEx-DFT}}\left[ v \right] = F^S\left[ \rho_0^S\left[ v_{eff} \right] \right]
+ E_{xc}^{sx}\left[ \rho_0^S\left[ v_{eff} \right] \right]
\nonumber
\\
+ \int dr v(r)  \rho_0^S\left[ v_{eff} \right] 
\end{eqnarray}

Now, the relation to Screened Exchange DMFT is becoming
clear: one may construct a spectral density functional in the
same spirit as in DFT+DMFT \cite{kotliar-rmp}, but starting from
the generalized KS functional.
In~\citen{amadon:066402}, the expression for the total energy within the
standard DFT+DMFT case was derived to be
\begin{equation}
E = E_{{\rm DFT}}-\sum_l \epsilon^{{\rm KS}}_l
+\bra H_{{\rm KS}}\ket+\bra( H_{int} - H_{dc}) \ket
\label{eq:energy_lda+dmft_1}
\end{equation}
where $\sum_l \epsilon^{{\rm KS}}_l$ is the sum of the occupied 
Kohn-Sham eigenvalues, $\bra H_{{\rm KS}}\ket =\t[H_{KS}\hat{G}]$,
and $H_{int}$ and $H_{dc}$ denote the local interaction part
of the Hamiltonian and the corresponding double counting term,
respectively.

Instead of using the usual Kohn-Sham Hamiltonian for
the construction of the one-body part, Screened Exchange
DMFT relies on the generalized Kohn-Sham reference system 
that includes the screened exchange potential.
The generalization of (\ref{eq:energy_lda+dmft_1})
to the present case thus replaces the
Kohn-Sham Hamiltonian $ H_{{\rm KS}}$ in the expression for
the energy by its non-local form,
keeping track of the effective potential part:
\begin{eqnarray}
E = E_{{\rm SEx-DFT}}-\sum_l \epsilon^{{\rm SEx-KS}}_l
+\bra \hat{\mathcal{O}} + \hat{v}_{eff} \ket
\nonumber 
\\
+\bra( H_{int}(V_{ee}, \lambda_s, \omega_s) - H_{dc}) \ket
\end{eqnarray}
Furthermore, the local interaction term is taken in the more
general form of a dynamical interaction, thus corresponding to
a local Hubbard term with {\it unscreened} interactions and local
Einstein plasmons of energy $\omega_s$ coupling to the electrons 
via coupling strength $\lambda_s$.

\begin{figure}
\includegraphics[scale=0.5]{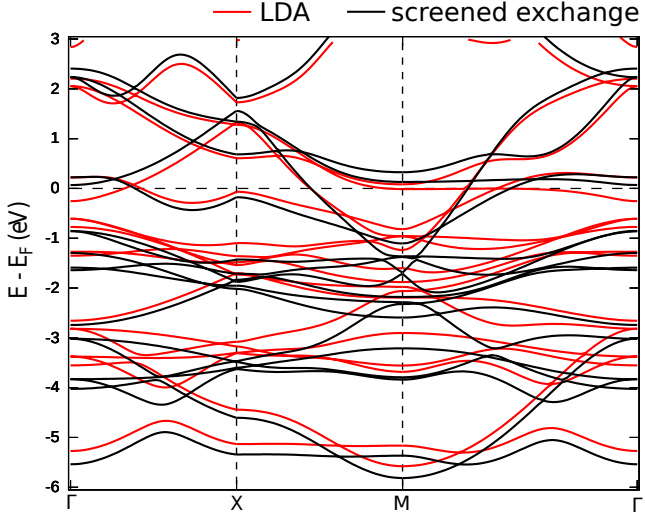}
\caption{(Color online) Kohn-Sham band structure of BaCo$_2$As$_2$ within DFT-LDA
(red lines) and the screened exchange approximation (black lines).}
\label{bands_LDA_SEX} 
\end{figure}

This concludes our description of the generalized Kohn-Sham
interpretation of Screened Exchange DMFT, resulting in particular in an
energy functional expression. However, in the practical calculations
presented in the following we do not 
minimise the full energy expression as given above, but rather work at the
converged DFT density and then investigate spectral properties
using the Screened Exchange DMFT formalism. This amounts to a 
one-shot Screened Exchange-DFT+DMFT calculation that uses 
the DFT density as a starting point.
The advantage of such an approach is obvious:
Numerically, this procedure allows us to avoid the 
expensive evaluation of non-local exchange terms within 
the self-consistency cycle of GKS theory. Moreover,
as is well-known, while severe deviations
of the true spectrum from the Kohn-Sham spectrum are quite common,
the ground state density obtained even from approximate DFT
functionals is often a good representation of the true one.
In the case of the {\it exact} DFT functional, our approach
would also lead to the exact ground state density and energy,
with additional improvements of the spectrum over standard 
Kohn-Sham DFT.


\begin{figure}
\includegraphics[width=0.5\textwidth]{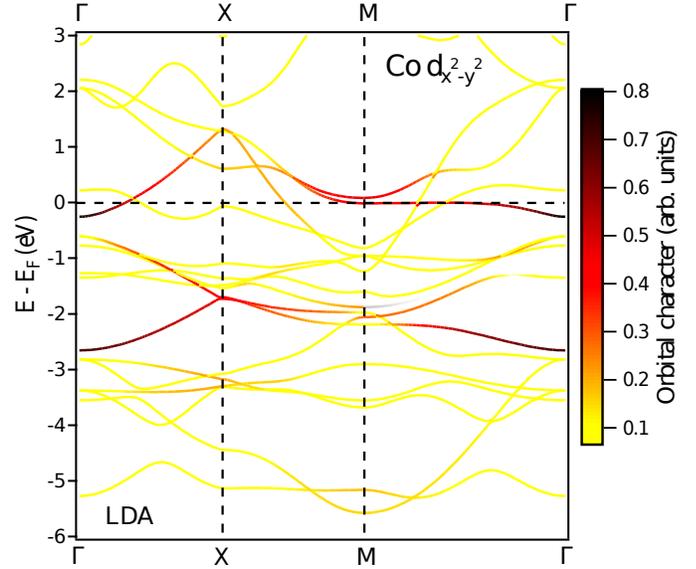}
\caption{(Color online) Kohn-Sham band structure of BaCo$_2$As$_2$ within DFT-LDA, 
projected on the $d_{x^2-y^2}$ orbital.}
\label{bands_dx2y2}
\end{figure}

\section{Results on BaCo$_2$As$_2$}
\label{sec:baco2as2}

As a first illustration of Screened Exchange DMFT, we review 
calculations on BaCo$_{2}$As$_{2}$, which is the fully Co-substituted
representative of the so-called ``122'' family of the iron-based
superconductors, isostructural to the
prototypical parent compound BaFe$_{2}$As$_{2}$. 
The Fe $\rightarrow$ Co substitution has however important
consequences: the nominal filling of the $3d$ states changes
from d$^{6}$ to d$^{7}$, which strongly reduces the degree of Coulomb
correlations, making BaCo$_{2}$As$_{2}$ a moderately
correlated compound\cite{BaCo2As2-Nan,PhysRevLett.113.266403}.
Indeed, the power-law deviations from Fermi 
liquid behavior above an extremely low coherence temperature
discussed in the Fe-based compounds\cite{Werner12} are a
consequence of the d$^{6}$ configuration and
strong intra-atomic exchange interactions.
This is no longer the case in the cobalt pnictides, where 
angle-resolved photoemission spectroscopy (ARPES) identifies 
well-defined and long-lived quasiparticle excitations
with relatively weak mass renormalization.
Nevertheless, the DFT-LDA derived Fermi surface differs from 
experiment\cite{BaCo2As2-Nan, BaCo2As2-Dakha, PhysRevLett.113.266403,
Brouet-BaCo2As2}. 
Therefore, BaCo$_{2}$As$_{2}$ provides an ideal testing ground for new
approaches to spectroscopic properties.
We do not discuss here the details of yet another interesting question,
which is the absence of ferromagnetism despite
a high value of the DFT density of states at the Fermi level, suggestive
of Stoner ferromagnetism, but refer the reader to Ref.~\citen{PhysRevLett.113.266403},
where the solution to this puzzle was discussed in detail.

\begin{figure}
\includegraphics[width=0.5\textwidth]{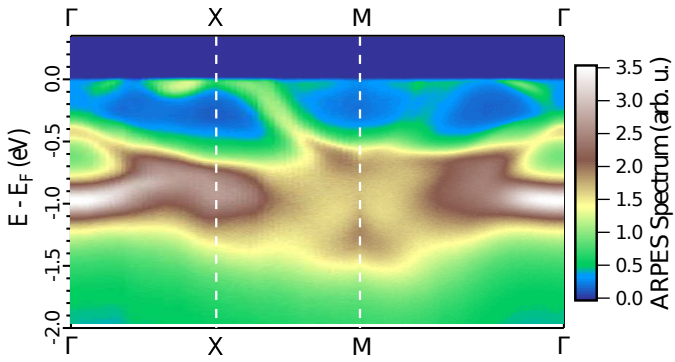}
\caption{(Color online) Angle-resolved photoemission spectrum of BaCo$_2$As$_2$.
Adapted from Ref.~\citen{PhysRevLett.113.266403}.
}
\label{BaCoAs_ARPES}
\end{figure}

\begin{figure}
\includegraphics[width=0.5\textwidth]{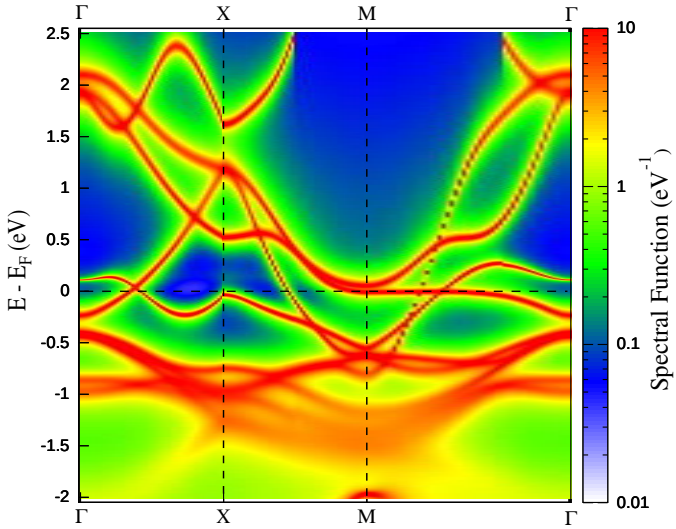}
\caption{(Color online) k-resolved spectral function of BaCo$_2$As$_2$  within LDA+DMFT.
Adapted from Ref.~\citen{PhysRevLett.113.266403}.}
\label{LDADMFT}
\end{figure}

\begin{figure}
\includegraphics[width=0.5\textwidth]{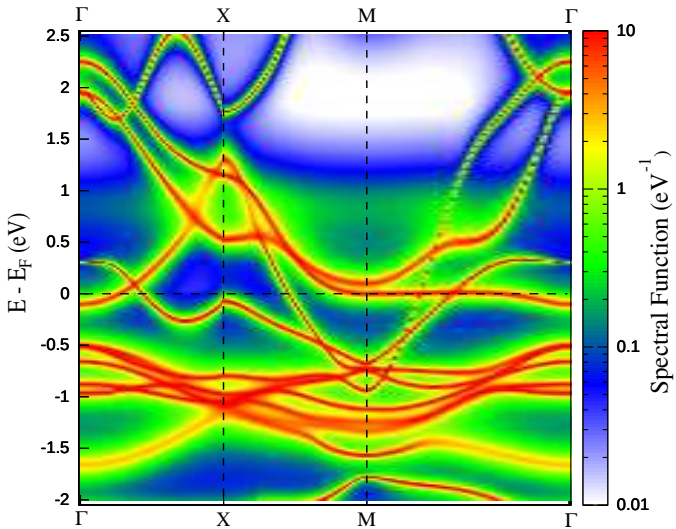}
\caption{(Color online) k-resolved spectral function of BaCo$_2$As$_2$  within 
Screened Exchange DMFT (with dynamical interactions). 
Adapted from Ref.~\citen{PhysRevLett.113.266403}.}
\label{SEXDMFT}
\end{figure}

\begin{figure}
\includegraphics[width=0.5\textwidth]{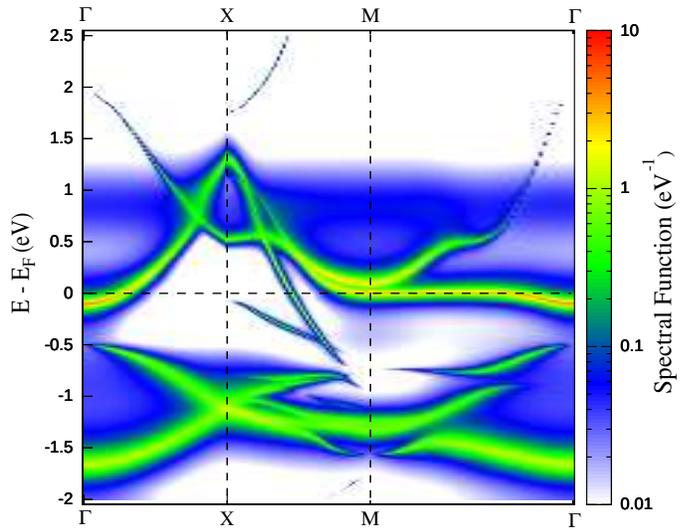}
\caption{(Color online) k-resolved spectral function of BaCo$_2$As$_2$ within 
Screened Exchange DMFT (with dynamical interactions), 
projected on the $d_{x^2-y^2}$ orbital.
Adapted from Ref.~\citen{PhysRevLett.113.266403}.}
\label{SEXDMFTx2y2}
\end{figure}

Fig.~\ref{bands_LDA_SEX} shows the DFT band structure of BaCo$_2$As$_2$,
in comparison to a screened exchange calculation.
As in the iron-based pnictides, the dominantly $3d$-derived states
are located around the Fermi level; in this case in a window of about -3~eV
to 2~eV. As compared to the parent iron pnictides with $3d^6$ configuration of the
Fe shell, the Fermi surface topology is modified due to the
larger $3d^7$ filling. The hole pocket at the $\Gamma$ point
that is present in most Fe-based pnictide compounds is pushed below the
Fermi level, as well as the band forming the electron pocket at $M$, 
which is now fully filled.
In standard DFT-LDA (see Fig.~\ref{bands_dx2y2}),
a characteristic flat band of dominant $x^2-y^2$ 
character lies directly on the Fermi level around the M point,
giving rise to a huge peak in the density of states. 
The close proximity of this band to the Fermi level will render its energy
highly sensitive to the details of the calculation
and thus provides a perfect benchmark for improved computational techniques.

Comparison to angle-resolved photoemission experiments\cite{PhysRevLett.113.266403},
as reproduced in Fig.~\ref{BaCoAs_ARPES}, reveals that the
overall bandwidth of the LDA band structure is too wide by roughly a factor of 1.5.
In a DFT+DMFT calculation, as shown in Fig.~\ref{LDADMFT}, 
this band renormalisation is reproduced, 
giving an overall occupied bandwidth of about 1.5~eV.
The fine details of the Fermi surface, and in particular the position 
of the flat $x^2-y^2$ band are however not well described.
For a detailed comparison we refer the reader
to Ref.~\citen{PhysRevLett.113.266403}.

Including screened exchange in the form of a Yukawa potential on top of DFT,
as shown in Fig.~\ref{bands_LDA_SEX} widens the band
by a considerable amount as compared to DFT-LDA. Even more striking
are the modifications at the Fermi level: the $x^2-y^2$ band
has been shifted above the Fermi level, with an energy at
the $M$ point of about 0.15~eV above $E_F$. While the lowering of the filling
of this band improves the agreement between theory and experiment,
the shift is too large to reproduce the experimental Fermi surface. 
However, when applying DMFT with dynamical interactions
on top of the screened exchange Hamiltonian, 
equivalent to the one-shot Screened Exchange DMFT procedure described above,
this band is renormalized by the electronic interactions and ends up again
close (but above!) the Fermi level. Its energy is slightly higher than in
DFT-LDA and is now in excellent agreement with the experimental Fermi
surface. A more detailed analysis presented in Ref. \cite{PhysRevLett.113.266403}
reveals that also the higher energy features such as the bands within
the range of up to 2~eV are well reproduced in this one-shot Screened 
Exchange DMFT approach.

The comparison of the band structures and spectral functions within the
different computational schemes illustrates the effects of the
different terms in an instructive way: Improving on the
description of exchange by replacing the local exchange
as contained in DFT-LDA by a non-local screened Fock exchange widens the bandwidth and 
strongly ``overcorrects'' the Fermi surface. Improving
at the same time on the correlation part by applying DMFT with frequency-dependent
interactions then leads to a partial cancellation of the band widening,
giving an overall bandwidth surprisingly close to the LDA+DMFT one.
This suggests that the good description of the overall bandwidth in
DFT+DMFT is the result of an error cancellation between 
the local approximation to exchange and partial neglect of correlations.
However, the Fermi surface is strongly modified by the non-local
corrections, leading to a significant improvement over LDA+DMFT and resulting 
in good agreement with experiment.

These findings might suggest that possible inconsistencies
between theoretical and experimental Fermi surfaces that are observed
in many pnictide materials could be corrected by adding simple
non-local self-energy corrections stemming from screened exchange
effects. 
Whether this treatment can fully account for the ``red-blue
shift'' of Fermi surface pockets found in the 
experimental-theoretical comparisons\cite{PhysRevLett.105.087001}
is a most interesting topic for future research.
The current example may give rise to optimism.


\section{How does Screened Exchange Dynamical Mean Field Theory
behave for weakly correlated materials?}
\label{sec:Zn_Cd}

\begin{figure}[t]
\begin{center}
\includegraphics[width=0.5\textwidth]{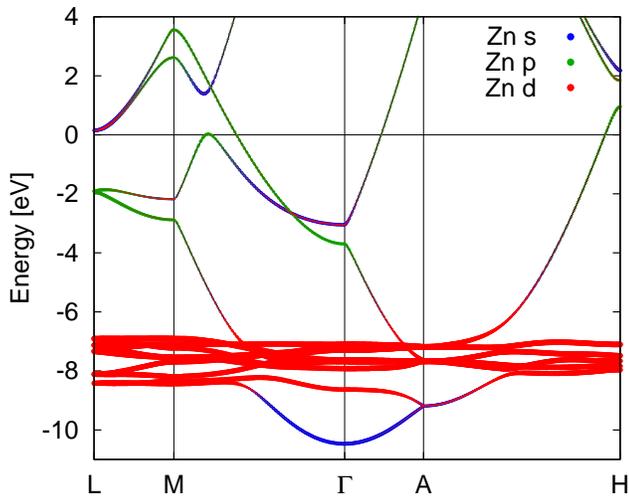}
\end{center}
\caption{(Color online) The band structure of elemental
Zn calculated within DFT. The orbital character is indicated
by the intensity of the different colors.
}
\label{fig:Zn_BS}
\end{figure}

\begin{figure}[t]
\begin{center}
\includegraphics[width=0.5\textwidth]{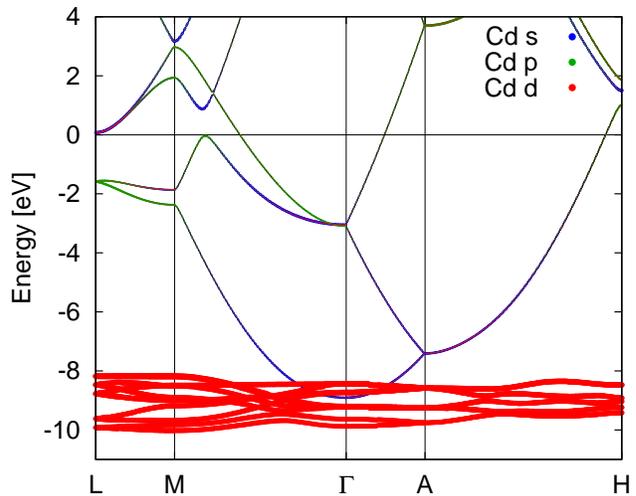}
\end{center}
\caption{(Color online) The band structure of elemental
Cd calculated within DFT. The orbital character is indicated
by the intensity of the different colors.
}
\label{fig:Cd_BS}
\end{figure}

Hamiltonians built as combinations of a DFT part
and local Hubbard-type interaction terms trivially reduce to the
DFT Kohn-Sham electronic structure when assuming that in weakly
correlated materials the effective local interactions become
negligible.
The question of the recovery of the weakly interacting limit is,
however, more interesting in the case of Screened Exchange DMFT.
While the static part of the effective local interaction may
be assumed to lose its importance, band widening
by the replacement of the DFT exchange correlation potential
by the non-local exchange-correlation GW self-energy persists. 
On the other hand,
plasmonic effects are also present in weakly correlated
materials and continue to renormalize the low-energy band
structure through electron-plasmon coupling.
This raises the question of what the resulting spectra for 
weakly correlated materials
look like in screened exchange + DMFT.

In Ref.~\citen{0295-5075-108-5-57003}, this question has been studied
for early transition metal perovskites, where it was found
that the band widening effect induced by non-local exchange and
the electronic polaron effect counteract each other and tend
to approximately cancel, thus resulting in a low-energy
electronic structure close to the DFT Kohn-Sham band structure
as long as static Hubbard interactions are disregarded.
Here, we address this question in the case of the seemingly
``simple'' transition metals zinc and cadmium.

Both elements nominally display a $d^{10}$ configuration, with
fully occupied $3d$ orbitals in the case of Zn and $4d$ in the case
of Cd, the dominantly $d$-derived bands being located several eV below the
Fermi level. In Fig.~\ref{fig:Zn_BS} and Fig.~\ref{fig:Cd_BS} we
show the band structure calculated within DFT for both materials.
Here and in the following we use the experimental crystal
structure.
DFT puts the occupied $d$ states at around -8~eV in Zn and
-9~eV in Cd. The conducting states of this 
transition metal are formed by dispersive 4$s$(5$s$) in Zn(Cd) states
around the Fermi level, that hybridize with the $p$-manifold.

These facts raise the immediate expectation of
negligibly small correlation effects on the occupied $d$ shells.
An effective Hubbard interaction calculated
for the $d$-manifold within the constrained random phase
approximation coincides with the fully screened
interaction, since as a consequence of the complete filling
of the d-shell there are no intra-d transitions to be
cut out as opposed to an open shell system, where transitions
inside the shell contribute to screening effects. 

\begin{figure}[t]
\begin{center}
\includegraphics[width=0.5\textwidth]{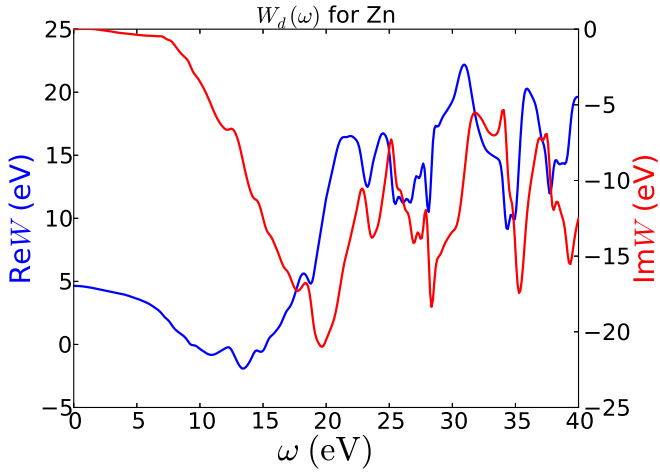}
\end{center}
\caption{(Color online) The fully screened effective local
Hubbard interaction on the $3d$ manifold for Zn.}
\label{fig:Zn_Ww}
\end{figure}

\begin{figure}[t]
\begin{center}
\includegraphics[width=0.5\textwidth]{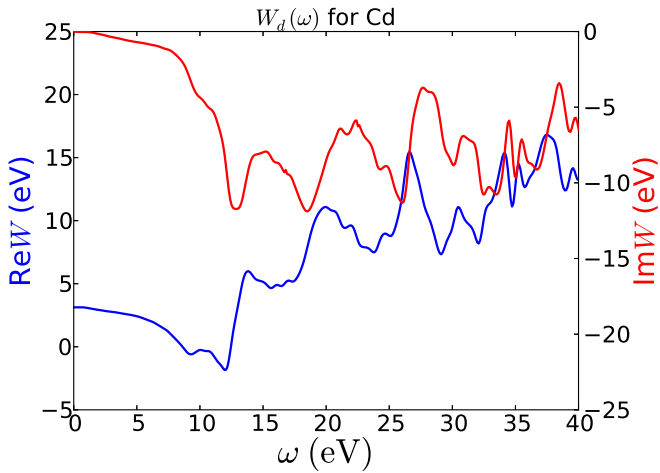}
\end{center}
\caption{(Color online) The fully screened effective local
Hubbard interaction on the $4d$ manifold for Cd.}
\label{fig:Cd_Ww}
\end{figure}

Figs.~\ref{fig:Zn_Ww} and \ref{fig:Cd_Ww} display the local component of this
fully screened interaction projected on the $d$-manifold.
The low-frequency
limit approaches a value of 4.8~eV and 3.1~eV for Zn
and Cd respectively. Even though this value is similar to
their oxides, which are open shell systems, where correlations on the $d$ states
are significant, the high binding energy of 
these states far away from the Fermi level effectively prevents
dynamical fluctuations. This suggests that screened exchange + DMFT
should in fact reduce to screened exchange renormalized by the bosonic
factor $Z_B$ discussed above. 
Nevertheless, this does not mean that static effects of the interaction
are properly treated in DFT. Even if this were the case,
there is no reason that the DFT Kohn-Sham spectrum, being
derived from an effective non-interacting system,
provides an accurate description of the experimental situation.

\begin{figure}[t]
\includegraphics[width=0.5\textwidth]{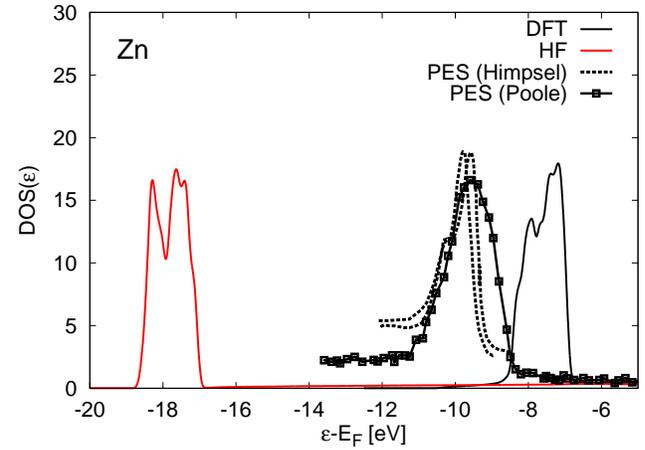}
\caption{(Color online) The Density of States of elemental Zn calculated within
Density Functional Theory (DFT, black solid line) and Hartree-Fock (HF, red solid line)
in comparison with Photoemission experiments~\cite{poole1973,himpsel1980} (dashed line, symbols).
Density Functional Theory (DFT) calculations underestimate
the binding energy of the Zn $3d$ states, while the HF overestimates the binding energy
significantly (see explanation in the text).}
\label{fig:DOS_Zn_HF}
\end{figure}

\begin{figure}[t]
\includegraphics[width=0.5\textwidth]{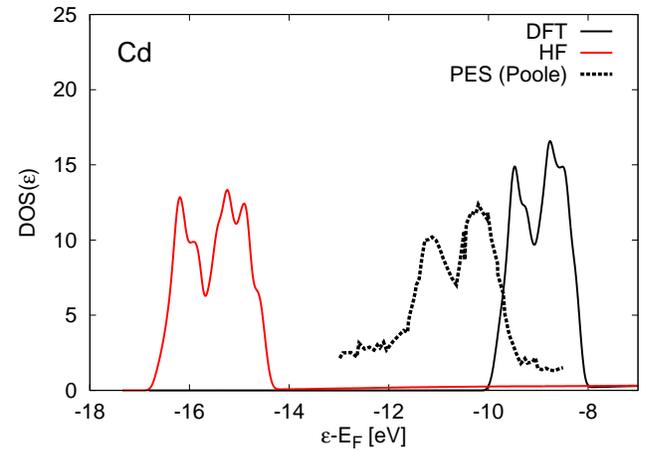}
\caption{(Color online) The Density of States of elemental Cd calculated within
Density Functional Theory (DFT, black solid line) and Hartree-Fock (HF, red solid line)
in comparison with Photoemission experiments~\cite{poole1973} (dashed line, symbols).
Density Functional Theory (DFT) calculations underestimate
the binding energy of the Cd $3d$ states, while the HF overestimates the binding energy
significantly (see explanation in the text).}
\label{fig:DOS_Cd_HF}
\end{figure}

Figs.~\ref{fig:DOS_Zn_HF} and \ref{fig:DOS_Cd_HF}
compare the experimental photoemission spectra~\cite{poole1973,himpsel1980}
from the literature to the density of states (DOS) derived
from DFT and Hartree-Fock (HF) theory.
The resulting discrepancy in terms of an underestimation
of the binding energy of the $d$ states in DFT of several eV
had been noted in the literature before\cite{norman1984,PhysRevB.54.17564,PhysRevB.60.10754}:
Norman \textit{et a.l} \cite{norman1984} discussed 
it in terms of a self-interaction error, proposing a correction 
in terms of an approximate substraction of self-interaction 
contributions contained in DFT~\cite{SIC_Perdew}.
Hartree-Fock calculations are, on the one hand, self-interaction free, but
on the other hand -- due
to the absence of screening -- widen all bands and  place the $d$-bands far too
low in energy, as can be seen in Figs.~\ref{fig:DOS_Zn_HF} and \ref{fig:DOS_Cd_HF}.

Since local dynamical correlations can assumed to be small
as discussed before, an improved treatment of the screened interaction
\textit{and} the self-interaction correction at the same time
is likely to improve the shortcomings of Hartree-Fock, resp. DFT.
Here we will discuss the two possible extensions of
Screened Exchange 
plus a bosonic renormalization factor $Z_B$ and the GW approximation.
The computationally cheaper option of screened exchange
is including only static exchange contributions with a
Yukawa-type interaction potential, and an effective renormalization $Z_B$
which originates from the spectral weight transfer to plasmonic 
excitations. GW is computationally more demanding, but has the advantage of
treating the dynamical part of the screening and correlation.
Even though both methods include a self-interaction correction,
the self-interaction contained in the Hartree term is not completely cancelled 
since the exchange contributions are derived from a screened interaction
and not the bare one, as opposed to the Hartree term. 

\begin{figure}[t]
\includegraphics[width=0.5\textwidth]{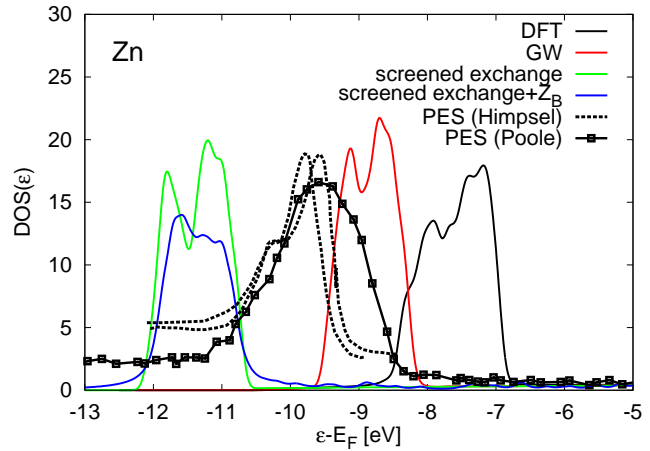}
\caption{(Color online) The Density of States of elemental Zn calculated within
different theoretical methods (solid lines), in comparison
with Photoemission experiments~\cite{poole1973,himpsel1980} (dashed line, symbols).
Density Functional Theory (DFT) calculations significantly underestimate
the binding energy of the Zn $3d$ states, while the GW approximation obtains 
a much better agreement. Screened exchange
overestimates the binding energy significantly.}
\label{fig:DOS_Zn}
\end{figure}

\begin{figure}[t]
\includegraphics[width=0.5\textwidth]{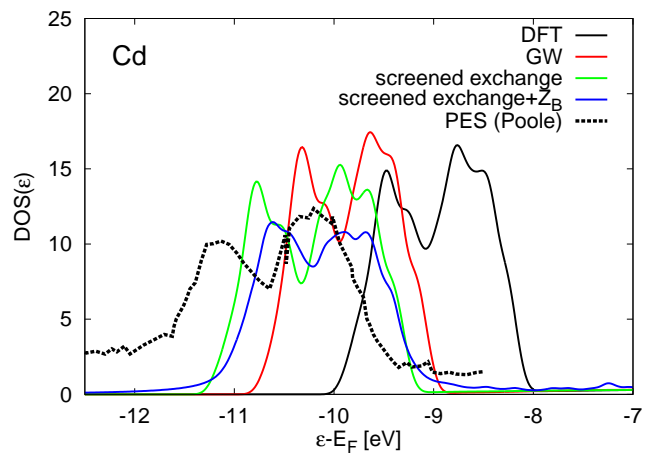}
\caption{(Color online) The Density of States of elemental Cd calculated within
different theoretical methods (solid lines), in comparison
with Photoemission experiments~\cite{poole1973} (dashed line).
DFT calculations significantly underestimate
the binding energy of the Cd $4d$ states, while the GW approximation
and also screened exchange obtain a much better agreement.}
\label{fig:DOS_Cd}
\end{figure}

In Figs.~\ref{fig:DOS_Zn} and \ref{fig:DOS_Cd}
we show comparisons of the DOS of zinc and cadmium, calculated within
DFT, screened Exchange(+$Z_B$) and GW\cite{jiang2013fhi} to  photoemission spectra.
In the GW calculation we used $7\times7\times3$ k-points and 5 additional
high-energy local orbitals.
Interestingly, while in both systems the GW approximation provides a significant
correction of the DFT Kohn-Sham spectrum in the right direction,
but still underestimates the binding energy, the screened exchange 
scheme places the d-states
too low in energy for Zn while providing a slightly better estimate
than GW in Cd. The addition of the bosonic
renormalization factor $Z_B$ merely renormalizes the d bandwidth
but keeps the average level position constant. 

This raises the interesting question
of which effects are missing in screened exchange and GW?
The incomplete cancellation of the self-interaction in both
approaches is expected to lead to an overall underestimation of the binding energy,
since the additional unphysical interaction increases the energy of
the d states. A more accurate estimate of 
this term would lead to an improvement of GW
in both systems, but an even larger error of screened exchange 
in Zn.
Another effect neglected in screened exchange is the Coulomb hole contribution: 
this term,
discussed by Hedin as part of the ``Coulomb hole screened
exchange (COHSEX)'' approximation translates the fact that
the presence of an electron at a position $r$ pushes away
charge at $r$ (in the language of a lattice model, the 
charge-charge correlation function exhibits a reduction 
of the double occupancy), and the interaction of this effective
positive charge with the electron presents an energy gain
expressed in the form of an interaction of the electron with
a ``Coulomb hole''.
This term, contained in GW but not in screened exchange, also increases the binding
energy. 

This leads to the overall picture that screened exchange with the 
inclusion of the static corrections just discussed has a tendency to overestimate
the binding energy in general, while GW underestimates it. 
The obvious difference between the two methods is the dynamical
treatment of the screened interaction, which is treated more appropriately
in GW, but it is not clear \textit{a priori}
whether the \textit{static} approximation of the screened interaction
or the approximated \textit{form} of the screened interaction
in terms of a Yukawa potential gives rise to the difference
between screened exchange and GW. 

The GW description of zinc and cadmium is close to the experiment.
The remaining discrepancy to experiment is likely
explained by remaining self-interaction contributions and/or missing self-consistency.
Self-consistency (or quasiparticle self-consistency) has been investigated in the homogeneous
electron gas~\cite{HolmSelfCGW} and various solid state 
systems~\cite{SchilfgardeQSGW}.
These questions are left for future work.

\section{Conclusions}
\label{sec:conclusions}
In this work we have reviewed and applied existing as well as novel
approaches to obtain spectral properties of correlated electron materials.
Guided by the need of a proper treatment of the long-range Coulomb
interaction and non-local exchange effects we presented a lightweight
version of the general GW+DMFT approach, the so-called Screened Exchange 
DMFT. It can be derived as a simplication to GW+DMFT
in terms of a generalized screened exchange DFT scheme 
where local interactions are treated by dynamical DMFT. 

Analysis and the application of a simplifed form of this scheme to 
BaCo$_2$As$_2$ 
indeed showed that non-local exchange and electronic screening
lead to significant corrections to the electronic spectrum
which are necessary to obtain a proper description of the experimental 
observations. 

Furthermore, we discussed the case of the elemental transition
metals Zn and Cd, where strong local correlations are unimportant but
the position of the occupied d manifold is 
very sensitive to a proper treatment of screened exchange effects.


\begin{acknowledgment}

SB and AvR thank their collaborators of Refs.36 and 37 for the
fruitful collaborations and discussions.
This work was supported by IDRIS/GENCI Orsay under project
t2017091393, the European Research Council under its
Consolidator Grant scheme (project 617196), and the ECOS-Sud grant A13E04.
HJ acknowledges financial support from National Natural Science Foundation 
(21373017, 21621061) and the
National Basic Research Program of China (2013CB933400).

\end{acknowledgment}


\end{document}